\documentclass[12pt]{article}
\usepackage{geometry} 
\usepackage{indentfirst}
\usepackage{setspace}
\usepackage[hidelinks]{hyperref}
\usepackage{amsmath,amsfonts,amsthm,amssymb}

\usepackage{graphicx}
\usepackage{mathtools}
\usepackage{xcolor}
\usepackage{float}
\usepackage{natbib, comment}

\usepackage{booktabs, tabularx, threeparttable, caption}
\numberwithin{figure}{section}
\numberwithin{table}{section}
\numberwithin{equation}{section} 
\makeatletter
\@addtoreset{equation}{section}
\makeatother

\title{Multiple Imputation for Nonresponse in Surveys Using Design Weights and Auxiliary Margins}

\author{Kewei Xu\thanks{Department of Statistical Science, Box 90251, Duke University, Durham, NC 27708-0251} \and Jerome P. Reiter\thanks{Department of Statistical Science, Box 90251, Duke University, Durham, NC 27708-0251}}
\date{}

\begin{document}
\maketitle

\begin{center}
    \textbf{Abstract}
\end{center}
\noindent

Survey data typically have missing values due to unit and item nonresponse. Sometimes, survey organizations know the marginal distributions of certain categorical variables in the target population. As shown in previous work, survey organizations can leverage these distributions in multiple imputation  for nonignorable unit nonresponse, generating imputations that result in plausible completed-data estimates for the variables with known margins.  However, this prior work does not use the design weights for unit nonrespondents.
We extend this previous work to utilize the design weights for all sampled units.  We illustrate the approach using simulation studies.

Key Words: Item; Missing; Nonignorable; Unit

\section{Introduction}

Survey data usually suffer from both unit and item nonresponse. As a result, survey organizations have to make strong assumptions about the reasons for missingness, for example,  the data are missing at random.
One way to lessen   reliance on assumptions is to utilize information in auxiliary data sources. For example, and pertinent to the setting of our work, 
survey organizations may know the population percentages or totals of some categorical variables in the survey.  These could be available from recent censuses, administrative databases, or other high quality surveys \citep{sadinle2017itemwise}. Indeed, such information is frequently used in methods for handling survey nonresponse, such as calibration and raking.

We consider settings where analysts seek to integrate auxiliary information on marginal distributions in multiple imputation for nonresponse \citep{rubin1987multiple} in surveys.  We build on the missing data with auxiliary margins, or MD-AM, framework introduced by \citet{akande2021leveraging} and extended to surveys by \citet{akande2021multiple}, \citet{tang2024using}, and \citet{yang:reiter}. The latter three works impute missing values to ensure completed-data, survey-weighted inferences are plausible given the known margins. However, these works impose a simplifying condition on the unit nonrespondents, namely that the analyst does not use their design weights and instead replaces these weights with a convenient constant.  

In this short note, we propose MD-AM models that allow use of the design-based weights for all sampled units. \textcolor{black}{These models are intended especially for unequal probability samples where one wants to incorporate relationships between the design variables and survey variables in imputation. They presume design weights are available for all sampled units, as should be true for the organization that collected the data.}
Section \ref{sec:review} provides a brief review of the hybrid missing MD-AM model \citep{tang2024using, yang:reiter}, which we extend to incorporate design weights for unit nonrespondents.  
Section \ref{sec:meth} introduces the methodology.  
Section \ref{sec:sims} illustrates the methods using simulation studies.  Codes and additional results are available at \url{https://github.com/kevinxu47/MDAM}.  

\section{Review of Hybrid Missingness MD-AM Modeling}\label{sec:review}

In reviewing the hybrid missingness MD-AM model, we closely follow the notation in \citet{yang:reiter}.  
 Let $\mathcal{P}$ be a finite population comprising $N$ units. For  $i = 1, \dots, N$, let $\boldsymbol{x}_i = (x_{i1}, \dots, x_{ip})$ denote the $p$ survey variables for unit $i$; 
 let $I_i = 1$ if unit $i$ is selected for inclusion in the survey and $I_i = 0$ otherwise; let $\pi_i = \Pr(I_i = 1)$ represent the probability that unit $i$ is selected into the survey; let $w_i^d = 1/\pi_i$  be its design weight; and, let $\boldsymbol{z}_i$ be its design variables, e.g., size measures or stratum indicators. Let $n = \sum_{i=1}^N I_i$ be the intended sample size. We refer to the sampled units as $\mathcal{S}$.


Let $U$ be the unit nonresponse indicator so that, for each unit $i \in \mathcal{S}$, $U_i=1$ if the unit provides no responses to any  survey questions and $U_i=0$ otherwise. 
Let \(\boldsymbol{R} = (R_1, \dots, R_p)\) be item nonresponse indicators corresponding to \(\boldsymbol{X} = (X_1, \dots, X_p)\) so that, for any unit \( i \) with \( U_i = 0 \) and any \( X_j \), \( R_{ij} = 1 \) if unit \( i \) does not respond to the question for \( X_j \) and \( R_{ij} = 0 \) otherwise. When $U_i=1$,  $(R_{i1}, \dots, R_{ij})$ are undefined. 

The survey organization has  auxiliary information $\mathcal{A}$ on the marginal distributions of a subset of $\boldsymbol{X}$. We write $X_j \in \mathcal{A}$ whenever $X_j$ has a known margin in $\mathcal{A}$, which we write as 
$\mathcal{A}_j$.
For convenience, we presume each $\mathcal{A}_j$ comprises population totals; incorporating percentages is a simple modification. 
For any categorical $X_j$ taking on levels $c=\{1, \dots, m_j\}$,
let $T_{jc} = \sum_{i=1}^NI(x_{ij}=c)$ be the total number of units in $\mathcal{P}$ at level $c$.  Here, $I(\cdot) = 1$ when the condition inside the parenthesis is true and $I(\cdot) = 0$ otherwise. 
 


\textcolor{black}{As described in \citet{tang2024using} and \citet{yang:reiter},} 
the hybrid missingness MD-AM model specifies a joint distribution of $(\boldsymbol{X}, \boldsymbol{R}, U)$.  
Specifically, the model presumes 
$U \sim \textrm{Bernoulli}(\pi_u)$,
where $\pi_u = \Pr(U=1)$ is the marginal probability of unit nonresponse. 
For  $j = 1, \dots, p$, let  $g_j(-)$ represent the model for $X_j$ given $X_1, \dots, X_{j-1}$, 
and let $\Omega_j$ and $\theta_j$ be the corresponding model parameters.  For example, $g_j(-)$ could be a logistic  regression of $X_j$ on some function of $X_1, \dots, X_{j-1}$, possibly including interactions, as well as a main effect for $U$ when $X_j \in \mathcal{A}$. 
Thus, for $\boldsymbol{X}|U$, the hybrid missingness MD-AM model uses  
\begin{eqnarray}
X_{1} \mid U &\sim& g_1(\Omega_{1}, \theta_{1}  U I(X_1 \in \mathcal{A}) \label{modelx1} \label{eq:x1an}\\
X_{j} \mid X_{1}, \dots, X_{j-1}, U &\sim& g_j(X_{1}, \dots, X_{j-1}, \Omega_j, \theta_{j} U I(X_j \in \mathcal{A}), \textrm{ for } j = 2, \dots, p.\label{eq:seqAN}
\end{eqnarray}
Because the  distribution for any $X_j \in \mathcal{A}$ differs for unit nonrespondents and respondents when  $\theta_j \neq 0$, the model encodes potentially missing not at random mechanisms for unit nonresponse.   The key identifying assumption is that the predictor function for any $X_j \in \mathcal{A}$ not include interactions between $U$ and elements of $(X_1, \dots, X_{j-1})$.  This assumption is a version of the additive nonignorable missingness mechanism  \citep{Hirano2001, sadinle2019sequentially}.  

For the model for each $R_j$, let 
$h_j(-)$ be some predictor function that excludes the corresponding $X_j$ but may include main and interaction effects involving other variables in $\boldsymbol{X}$. For $j=1, \dots, p$, we have    
\begin{equation}\label{Rmodel}
R_{j} \mid X_{1}, \dots, X_{p}, U=0 \sim \text{Bernoulli}(\pi_{R_{j}}),\,\,\, \text{logit}(\pi_{R_{j}}) =  h_j(X_{1}, \dots, X_{j-1}, X_{j+1}, \dots X_{p}, \Phi_j). 
\end{equation}
The models  in \eqref{Rmodel} encode itemwise conditionally independent nonresponse (ICIN) mechanisms \citep{sadinle2017itemwise}, which are known to have identifiable parameters. 

The imputation and estimation algorithms in \citet{tang2024using} and \citet{yang:reiter} do not use the design weights for unit nonrespondents.  Instead, they replace the design weights for unit nonrespondents with a constant that 
makes $\sum_{i=1}^N w_i = N$; that is, they base imputations and completed-data inferences on, for all units $i \in \mathcal{S},$  
\begin{equation}
w_i =  \left\{\begin{array}{lr}
        w_i^d & \text{if } U_i = 0\\
        \frac{N-\sum_{k \in \mathcal{S}} w_k^d}{\sum_{k \in \mathcal{S}} U_k} & \text{if } U_i = 1.
        \end{array}\right. \label{known_weights_unit_res}
\end{equation}


For any unit $i \in \mathcal{S}$ and $j=1, \dots, p$, let $x_{ij}^\star = x_{ij}$ when $(R_{ij} = 0, U_i=0)$, and let $x_{ij}^\star$ be an imputed value when $R_{ij} = 1$ or $U_i=1$.   For any $X_j \in \mathcal{A}$, 
\citet{akande2021multiple}, \citet{tang2024using}, and \citet{yang:reiter} suppose the imputations  adhere to  \eqref{modelx1} -- \eqref{Rmodel}, with the additional constraint that 
\begin{equation}
\hat{T}_{jc}^* = \sum_{i \in \mathcal{S}}w_i I(x_{ij}^{\star} = c) \sim N(T_{jc},V_{jc}),
\label{prob_cons}
\end{equation}
where the weights are from \eqref{known_weights_unit_res}.
Here, $V_{jc}$ is set by the analyst and determines how closely $\hat{T}_{jc}^*$ matches $T_{jc}$ in any completed dataset. 


\textcolor{black}{Using \eqref{prob_cons} for imputations has commonalities with   balanced random imputation \citep{chauvet:deville:haziza, chauvet:paco}, although the goals differ. In balanced random imputation, the imputed values are required always to satisfy some enforced constraint, e.g., the survey-weighted mean of the imputed values equals the survey-weighted mean of the observed values.  
Work on balanced random imputation tends to consider single imputations for item nonresponse, presume values are missing (completely) at random, and set constraints based on observed data quantities. In contrast, the hybrid missingness MD-AM model considers multiple imputation for unit nonresponse that is missing not at random, and it uses auxiliary marginal information to establish distributional constraints on imputations.} 

\textcolor{black}{Although not done in  \citet{tang2024using} or \citet{yang:reiter},  the model for each $X_j$ or $R_j$ can include the design variables $\boldsymbol{Z}$
as predictors.  Including $\boldsymbol{Z}$ can improve explanatory power, and hence ultimately imputation quality and estimation accuracy, when they are associated with the survey variables or nonresponse indicators \citep{kinney:reiter:raghu}.  Alternatively, the models can include some function of $W$ 
as a predictor. This is common advice for imputation modeling \citep[e.g., ][]{kimetal, quartagno} that is particularly salient when  
 the analyst doing the imputations has the survey weights but not the design variables, e.g., because they are  confidential. In the MD-AM models of \citet{tang2024using} and \citet{yang:reiter}, however, every unit nonrespondent has the same value of $w_i$ given in \eqref{known_weights_unit_res}.  Thus, including $W$ in the models would not influence the imputation probabilities, and hence not contribute to incorporating the design in the imputation process, for the unit nonrespondents.}

\section{Methodology}\label{sec:meth}

We now modify 
the hybrid missingness MD-AM model 
to use the design weights for all sampled individuals, including unit nonrespondents. \textcolor{black}{For brevity of notation, we now let $W$ refer to the design weights $w_i^d$ rather than the weights in \eqref{known_weights_unit_res}.   We follow the general  strategy outlined in \citet{yang:reiter}. First, we impute values for item nonresponse. Second, we use  $\mathcal{A}$  to impute values of variables with margins for unit nonrespondents.  Third, we impute values of the remaining variables for unit nonrespondents. Our innovation is to modify the second step so that imputations and the completed datasets use the design weights rather than \eqref{known_weights_unit_res} for  unit nonrespondents. This change requires new imputation algorithms, which we present in Section \ref{sec:estimation}.  For the first and third steps, we use techniques presented in \citet{yang:reiter}, which we briefly summarize in Section \ref{sec:mice} and Section \ref{sec:hotdeck}.}
For convenience, we presume that $X_j \in \mathcal{A}$ for $j = 1, \dots, k<p$, and the remaining $p-k$ variables do not have auxiliary margins.


\subsection{Variables with Item Nonresponse}\label{sec:mice}

We first create multiple imputations for all values of  $x_{ij}$ missing due to item nonresponse for units with $U_i=0$.  
We implement multiple imputation by chained equations (MICE)  
using only units with $U_i=0$ and potentially including $\boldsymbol{Z}$ as predictors.  
The specification of the MICE algorithm does not require any extraordinary considerations.
In our simulations,  we use the ``mice'' package in R \citep{vanbuuren}, 
following its default settings including the ordering of variables.
We run the MICE procedure to  create $L$ completed datasets for the units with $U_i=0$.

\subsection{Variables with Margins for Unit Nonrespondents}\label{sec:estimation}

After imputing values for item nonresponse, in each completed dataset we impute the unit nonrespondents' values for all $X_j \in \mathcal{A}$.  The analyst orders these variables based on ease of modeling. For  convenience, we suppose that $X_1$ is imputed first,  
 $X_2$ is imputed second, and so on sequentially until $X_k.$

To begin, for each unit $i$ with $U_i=1$ and in each of the $L$ completed datasets, the analyst sets an initial probability distribution for imputing $x_{i1}$.  For example, in each completed dataset, 
the analyst estimates a (multinomial) logistic regression of $X_1$ on \textcolor{black}{some function of $\boldsymbol{Z}$ (or $W$),}   
and uses 
the predicted probabilities from the estimated model.   
Alternatively, the analyst  could regress $X_1$ on an intercept only or compute marginal probabilities of $X_1$ using survey-weighted ratio estimators. For these two alternatives, the initial  probabilities are identical for all unit nonrespondents.  This approach may be especially appropriate when the analyst does not expect associations between the design weights and the missing values of $X_1$ for unit nonrespondents.



Regardless of how it is derived, we refer to the initial distribution for imputing $x_{i1}$ as its working distribution, notated as $\{p_{i1c}: c = 1, \dots, m_1; U_i=1; i \in \mathcal{S}\}.$ Because of imputation for item nonresponse, the working  distribution for any unit can differ across completed datasets; however, for convenience, we forego notation designating completed datasets. The analyst performs the computations of this section for each of the $L$ completed datasets.

Imputing each missing $x_{i1}$ using its working distribution does not utilize the information in $\mathcal{A}_1$.  The resulting imputations could generate completed-data \citet{horvitz1952generalization} estimates that are far from the known totals for some levels of $X_1$, \textcolor{black}{particularly when values are missing not at random.} We therefore modify  $\{p_{i1c}\}$ to ensure the resulting imputations result in reasonable design-based, completed-data estimates of these totals. \textcolor{black}{This is the primary objective and motivation of our extension of the MD-AM framework.}



To do so, 
we first simulate plausible values of $\hat{T}_{1c}$ for each level $c$ by appealing to large-sample central limit theorems. For $c=1, \dots, (m_1-1)$, we sample  $\hat{T}_{1c}$ from  $N(T_{1c}, V_{1c})$ and set $\hat{T}_{1m_1} = N - \sum_{c=1}^{m_1-1} \hat{T}_{1c}$. 
We then find adjusted probabilities $\{\tilde{p}_{i1c}: c=1, \dots, m_1; U_i=1;  i \in \mathcal{S}\}$ so that, for  $c = 1, \dots, m_1$, the expectation of each completed-data estimator $\hat{T}_{1c}^{\star}$ from \eqref{prob_cons} 
over imputations for unit nonresponse approximately equals $\hat{T}_{1c}$.  Put another way, recognizing that 
\begin{eqnarray*}
\mathbb{E} \left[\sum_{i \in \mathcal{S}}w_i I(x_{ij}^{\star} = c)\right] 
&=& \sum_{i \in \mathcal{S}} w_i I(x_{ij}^{\star} = c)I(U_i=0) + \mathbb{E} \left[\sum_{i \in \mathcal{S}}w_i I(x_{ij}^{\star} = c)I(U_i=1) \right] \nonumber\\
&=& \sum_{i \in \mathcal{S}} w_i I(x_{ij}^{\star} = c)I(U_i=0) + \sum_{i \in \mathcal{S}}w_i \tilde{p}_{i1c}I(U_i=1),
\end{eqnarray*}
we want  
\begin{eqnarray*}
\sum_{i \in \mathcal{S}}w_i \tilde{p}_{i1c}I(U_i=1) 
&=&  \hat{T}_{1c} - \sum_{i \in \mathcal{S}} w_i I(x_{i1}^* = c)I(U_i = 0).  \label{eq:modified}
\end{eqnarray*}


To keep the imputation probabilities tied to the working distribution, for $c=1, \dots, m_1-1$, we set $\tilde{p}_{i1c} = f_{1c} p_{i1c}$, where each constant $f_{1c}>0$ is given by   
\begin{equation}
f_{1c}=   \frac{\hat{T}_{1c} - \sum_{i \in \mathcal{S} } w_i I(x_{i1}^* = c, U_i = 0)}{\sum_{i \in \mathcal{S}} p_{i1c} w_i I(U_i = 1)}.  \label{eq:f1}
\end{equation}
We impute  $x_{i1}^*$ for each unit with $U_i=1$ by drawing randomly from the adjusted probability distribution, $\{\tilde{p}_{i1c}: c = 1, \dots, m_1\}$, where 
$\tilde{p}_{i1m_1} = 1 - \sum_{c=1}^{m_1-1} f_{1c}{p}_{i1c}$. If $\sum_{c=1}^{m_1-1} f_{1c}{p}_{i1c}>1$, we compute \eqref{eq:f1} for $c=1, \dots, m_1$ and set  $\tilde{p}_{i1c}= f_{1c}{p}_{i1c} / \sum_{c=1}^{m_1}f_{1c}{p}_{i1c}$. In the event that some  $f_{1c}p_{i1c} < 0$, 
we set those $\tilde{p}_{i1c} = 0$. Analysts may consider reducing $V_{1c}$ in this case.

We now turn to imputing $X_2 \in \mathcal{A}$ for unit nonrespondents. We follow a similar strategy: in each completed dataset, start by setting each unit nonrespondent's working distribution for imputation of $x_{i2}^*$ given $x_{i1}^*$ and $(\boldsymbol{z}_i, w_i$), then adjust the probabilities to make $(\hat{T}_{21}^*, \dots, \hat{T}_{2m_2}^*)$ approximately match a sampled value of $(\hat{T}_{21}, \dots, \hat{T}_{2m_2}).$  \textcolor{black}{We present two options for imputing $X_2$, one based on applying a multiplicative adjustment to the working probabilities (Section \ref{sec:mdamadj}) and the other based on solving a system of equations making use of $\mathcal{A}$ (Section \ref{sec:mdamsys}).}

\subsubsection{Multiplicative Adjustment Method}\label{sec:mdamadj}

Let $p_{i2c} = \Pr(X_{i2} = c | x_{i1}^*=d, \boldsymbol{z}_i, w_i)$ be the working probability that $X_2=c$ for unit $i$, given its imputed $x_{i1}^*=d$ and $(\boldsymbol{z}_i, w_i)$.  
We determine these from a (multinomial) logistic regression of $X_2$ on $X_1$ and possibly some function of $\boldsymbol{Z}$ or $W$, with coefficients estimated from the completed data for units with $U_i=0$. Note that if we disregard $\boldsymbol{Z}$ and $W$, $p_{i2c} 
= \Pr(X_{2} = c | X_{1}=d)$ for $c=1, \dots, m_2$ and $d=1, \dots, m_1$; that is, the conditional probability 
is the same for all units with $x_{i1}^*=d$.  For this case (which we use in Section \ref{sec:mdamsys}), to simplify notation we drop the index $i$ for individual units and add a subscript $d$ for the value of $X_1$, writing  $p_{i2c} = p_{2cd}$.

\textcolor{black}{
For $c=1, \dots, m_2-1$, we sample a plausible value of $\hat{T}_{2c}$ from a $N(T_{2c}, V_{2c})$, setting the estimated total for the final level as $\hat{T}_{2m_2} = N-\sum_{c=1}^{m_2-1}\hat{T}_{2c}$.  As in \eqref{eq:f1}, we find  suitable constants $f_{2c}>0$ to adjust each $p_{i2c}$. We have 
\begin{equation}
f_{2c}=   \frac{\hat{T}_{2c} - \sum_{i \in \mathcal{S} } w_i I(x_{i2}^* = c, U_i = 0)}{\sum_{i \in \mathcal{S}} p_{i2c} w_i I(U_i = 1)}.  \label{eq:f2}
\end{equation}
We impute each $x_{i2}^*$ given 
$x_{i1}^*$ and $\boldsymbol{z}_i$ (or $w_i$) for units with $U_i=1$ using a random draw from the adjusted probability mass function $\{\tilde{p}_{i2c}: c = 1, \dots, m_2\}$, where   $\tilde{p}_{i2c}= f_{2c}{p}_{i2c}$ for $c=1, \dots, m_2 -1$ and $\tilde{p}_{i2m_2} = 1 - \sum_{c=1}^{m_2-1} f_{2c}{p}_{i2c}$.  If $\sum_{c=1}^{m_2-1} f_{2c}{p}_{i2c}>1$, we compute \eqref{eq:f2} for $c=1, \dots, m_2$ and  set
$\tilde{p}_{i2c}= f_{2c}{p}_{i2c} / \sum_{c=1}^{m_2}f_{2c}{p}_{i2c}$. As for $X_1$, if some 
$f_{2c}{p}_{i2c} < 0$, we set those $\tilde{p}_{i2c}=0$. This process ensures the completed-data \citet{horvitz1952generalization} estimates approximately match the sampled $(\hat{T}_{21}, \dots, \hat{T}_{2m_2})$ in expectation, while also incorporating relationships between $X_2$, $X_1$ and $\boldsymbol{Z}$ (or $W$) implied in the working probabilities.
We independently repeat the  estimation and imputation process in each completed dataset.
}  \textcolor{black}{We refer to this method as MDAM--adj.}


When  $\{p_{i2c}\}$
derives from a logistic regression of $X_2$ on $X_1$ and $\boldsymbol{Z}$ (or $W$), using $\{\tilde{p}_{i2c}\}$ is equivalent to adjusting the intercept in that regression, 
leaving other coefficients alone.  This implies a model in which the log-odds for $X_1$ are the same for unit respondents and  nonrespondents, 
in accordance with 
no interactions between $U$ and $X_1$ in  \eqref{eq:seqAN}. 

\subsubsection{Systems of Equations Method} \label{sec:mdamsys}

MDAM--adj captures the relationship between $X_2$ and $X_1$ through the logistic regression used in the working probabilities.  Another approach is to impose constraints on the imputations through a system of equations implied by the assumptions underpinning  
\eqref{eq:seqAN} and \eqref{prob_cons}, as we now describe.

As in Section \ref{sec:mdamadj}, we first sample plausible values of $(\hat{T}_{21}, \dots, \hat{T}_{2m_2})$.  For now, we presume the working probabilities follow $p_{i2c} = p_{2cd}$ for all units.  
\textcolor{black}{
Within any completed dataset, 
we encode the constant log-odds assumption assumption in \eqref{eq:seqAN} 
as a set of $(m_1-1)(m_2 - 1)$ equations.  For  $c=2, \dots, m_2$ and  $d=2, \dots, m_1$, we have }
\begin{equation}
 \text{log} \left[\frac{p_{2cd}}{p_{21d}}\right] - \text{log} \left[\frac{p_{2c1}}{p_{211}} \right] 
= \text{log}\left[\frac{\Pr(X_2 = c| X_1 = d, U=0)}{\Pr(X_2=1| X_1=d, U=0)}\right] - \text{log} \left[\frac{\Pr(X_2=c| X_1=1, U=0)}{\Pr(X_2=1| X_1=1, U=0)} \right]. \label{eq:X2logodds} 
\end{equation}

We estimate the conditional probabilities for unit respondents in \eqref{eq:X2logodds} via survey-weighted ratio estimators using the  completed data; for example, 
\begin{equation}
\Pr(X_2 = c | X_1 = d, U=0) = \frac{ \sum_{i \in \textit{S} } w_i  I(x_{i1}^* = d, x_{i2}^* = c, U_i = 0) }{\sum_{i \in \textit{S} } w_i  I(x_{i1}^* = d, U_i = 0)}. \label{eq:survwtprob}
\end{equation}

We encode the conditions  on 
$(\hat{T}_{21}^*, \dots, \hat{T}_{2m_2}^*)$ 
in \eqref{prob_cons} as a set of $m_2$ equations.  We have 
\begin{equation}
\sum_{d=1}^{m_1} \sum_{i \in \mathcal{S}}  w_i {p}_{2cd} I(x_{i1}^* = d, x_{i2}^* = c, U_i = 1)  = \hat T_{2c} - \sum_{i \in \mathcal{S}} w_i I(x_{i2}^* = c, U_i = 0).  \label{eq:x2total}
\end{equation}

We solve these equations for the $m_1(m_2-1)$ conditional probabilities,
$\{p_{2cd}: c=2, \dots, m_2; d = 1, \dots m_1\}$. We  impute $x_{i2}^*$ given $x_{i1}^*$ for the unit nonrespondents using Bernoulli draws with probabilities $\{{p}_{2cd}\}.$  We refer to this method as MDAM--sys.

It is possible to modify MDAM--sys to use working probabilities $\{p_{i2c}\}$ that vary for units with the same value of $x_{i1}^*$.  To do so, 
we impute using Bernoulli draws with probabilities $\{\tilde {p}_{i2c} \}$, where $\tilde {p}_{i2c} =  f_{2cd} {p}_{i2c}$ and  
\begin{equation}
f_{2cd} = \frac{\sum_{i \in \mathcal{S}}w_i {p}_{2cd}}{\sum_{i \in \mathcal{S}}w_i {p}_{i2c}} \label{eq:sysadj}
\end{equation}
for each $(X_2 = c, X_1 = d)$.  The adjustment in \eqref{eq:sysadj} is motivated by matching 
$\sum_{i \in \mathcal{S}} w_i {p}_{2c} = \sum_{i \in \mathcal{S}}w_i f_{2cd} {p}_{i2c}.$

\subsubsection{Accounting for Additional Variables with Margins}
When $k>2$, for MDAM--adj 
we can apply the process used in \eqref{eq:f2} to each $X_j \in \mathcal{A}$.  For example, if $X_3 \in \mathcal{A}$, the analyst specifies a set of working probabilities, $p_{i3c} = \Pr(X_{i3}=c | x_{i1}^*, x_{i2}^*, \boldsymbol{z}_i,w_i)$ for $c=1, \dots, m_3$ via a logistic regression of $X_3$ on $X_1, X_2,$ and $\boldsymbol{Z}$ (or $W$).  The analyst samples values of $(\hat{T}_{31}, \dots, \hat{T}_{3m_3})$ and computes values of $f_{3c}$ using expressions analogous to \eqref{eq:f2}, replacing quantities based on $X_2$ with those based on $X_3$. 
For MDAM--sys, the analyst can solve a system of equations akin to those in \eqref{eq:X2logodds} and \eqref{eq:x2total}.  However,  MDAM--sys becomes increasingly complicated as more variables are added to $\mathcal{A}$.   

\subsection{Variables Without Margins for Unit Nonrespondents}\label{sec:hotdeck}


After creating $L$ partially completed datasets using  Section \ref{sec:mice} and \ref{sec:estimation}, we impute values  $x_{ij}^*$ for unit nonrespondents for all $X_j \notin \mathcal{A}$. We use the random hot deck imputation procedure developed by \citet{yang:reiter}. 
Paraphrasing from their presentation, 
for any unit $i \in \mathcal{S}$ with $U_{i}=0$, in any completed dataset, 
let $\boldsymbol{x}_{i}^{\mathcal{A}} = \{x_{ij}: X_j \in \mathcal{A}; j=1, \dots, p; U_i=0\}$ be the values of the (possibly imputed) survey variables in the completed data for those $X_j \in \mathcal{A}$.  
Similarly, for any unit $i' \in \mathcal{S}$ with $U_{i'}=1$, let  $\boldsymbol{x}_{i'}^{\mathcal{A}*} = \{x_{i'j}^*: X_j \in \mathcal{A}; j = 1, \dots, p; U_i=1\}.$ 
For each unit $i'$ with $U_{i'}=1$,  in each completed dataset we construct its donor set,  
$\mathcal{D}_{i'} = \{(x_{i1}, \dots, x_{ip}): \boldsymbol{x}_{i}^{\mathcal{A}}=\boldsymbol{x}_{i'}^{\mathcal{A}*}, U_{i}=0, i \in \mathcal{S}\}$. 
We  randomly sample one record $i$ from $\mathcal{D}_{i'}$ and append its  $\{(x_{i1}, \dots, x_{ik}): X_j \notin \mathcal{A}\}$ to $\boldsymbol{x}_{i'}^{\mathcal{A}*}$ to make the full  imputation $\boldsymbol{x}_{i'}^*$ for unit $i'$. We apply this procedure for all units in $\mathcal{S}$ with $U_{i'}=1$, resulting in a completed dataset for all $n$ units. We repeat this process in each of the $L$ datasets.

\section{Simulation Studies}\label{sec:sims}

We construct a population $\mathcal{P}$ comprising $N=3,405,809$ individuals from the 2023 American Community Survey public use files available from IPUMS. Each individual has the six variables described in Table \ref{tab:cps-variables}.  
We fill in any missing values using a single bespoke  run of the ``mice'' package. We treat the survey weight on the file as a known size measure, 
$Z= (z_1, \dots, z_N)$. To generate any $\mathcal{S}$, we sample records from $\mathcal{P}$ independently with probabilities $\pi_i=1/10z_i$,  where $i=1, \dots, N$. This Poisson sampling results in approximately \textcolor{black}{$n=7000$} individuals in any $\mathcal{S}$. We take 500 independent Poisson samples.

\begin{table}[t]
\centering
\begin{threeparttable}
\caption{Description of variables in  
the simulation study.}
\label{tab:cps-variables}
\begin{tabularx}{\textwidth}{@{} l c X c @{}}
\toprule
\textbf{Variable} & \textbf{Notation} & \textbf{Range} \\
\midrule
Sex                               & $X_1$ & 1 = Male, 0 = Female  \\
Marital status                    & $X_2$ & 1 = Married, 0 = Other \\
Any health insurance coverage     & $X_3$ & 1 =  Yes, 0 = No \\
School attendance                 & $X_4$ & 1 = Yes, 0 = No \\
Travel time to work               & $X_5$ & 0 to 888 \\
Occupational income score         & $X_6$ & 0 to 80 \\
\bottomrule
\end{tabularx}
\end{threeparttable}
\end{table}

  For each $\mathcal{S}$, we generate unit nonresponse by sampling $U_i$ for all $i \in \mathcal{S}$ from  
\begin{equation}
\label{1a}
U  \sim \text{Bernoulli}(\pi_{U})\quad \text{logit}(\pi_{U})=\omega_{0} + \omega_{1} X_1 + \omega_{2} X_2.
\end{equation}
We set $(\omega_{0}, \omega_{1}, \omega_{2}) = (-1.6, 0.5, 0.5)$ for a unit nonresponse rate around \textcolor{black}{25\%}. We make all data other than $Z$ and $W$ completely missing for every unit where \( U_i = 1 \). To generate item nonresponse for any unit $i$ with $U_i=0$, we sample  $R_{ij}$ for $j=1, \dots, 6$ using 
\begin{equation}
\label{2h}
\begin{aligned}
& R_{j} \mid X_1, X_2, X_3, X_4, X_5, X_6, Z, U=0 
   \sim \mathrm{Bernoulli}(\pi_{R_j}), \\
& \operatorname{logit}(\pi_{R_j}) 
   = \phi_{j0} + \sum_{k \neq j} \phi_{jk} X_k 
     + \phi_{Z} Z.
\end{aligned}
\end{equation}
 We set  parameters in \eqref{2h} so that approximately \textcolor{black}{20\%} of each survey variable is missing for units with $U_i=0$. Specifically, 
$\phi_Z = 0.00001$; 
\textcolor{black}{$\phi_{j0} = -1.5$} when $j \leq 4$ and \textcolor{black}{$\phi_{j0} = -1.6$} when $j =5,6$; and, $\phi_{jk} = 0.1$ when $k \leq 4$ and $\phi_{jk}=0.0001$ when $k=5,6$,  for all $j$.  We blank every $x_{ij}$ for which the sampled $R_{ij}=1$.

We let $\mathcal{A}$ include 
$T_{X_1}=\sum_{i=1}^N x_{i1}$ and $T_{X_2}=\sum_{i=1}^N x_{i2}$.
Using Section \ref{sec:mice}, we generate $L=10$ completed datasets for item nonresponse   using a bespoke implementation of the ``mice'' package. 
Using Section \ref{sec:estimation}, we 
implement MDAM--adj and MDAM--sys with working probabilities derived from logistic regressions of $X_1$ on $Z$ and of $X_2$ on $(X_1, Z)$. 
For $V_{X_1}$ and $V_{X_2}$ used to  
sample  $\hat{T}_{X_1} \sim N(T_{X_1}, V_{X_1})$ and $\hat{T}_{X_2} \sim N(T_{X_2}, V_{X_2})$,  
we use ``mice'' to make one completed dataset for all of $\mathcal{S}$ and compute the expressions for the unbiased variance estimators for $\hat{T}_{X_1}$ and $\hat{T}_{X_2}$ under Poisson sampling.  Using Section \ref{sec:hotdeck},  we create donor pools for the hot deck  by matching on $(X_1, X_2)$.

After generating $L=10$ completed datasets, we use the inferential methods in \citet{rubin1987multiple} for  the 
marginal totals for $(X_1, \dots, X_6)$ and several conditional and joint probabilities; see 
\url{https://github.com/kevinxu47/MDAM} for the full list.  
In each completed dataset, we compute \citet{horvitz1952generalization} point and variance estimators under Poisson sampling based on the design weights for all sampled records.
For comparison, we also use the method in \citet{yang:reiter} with the weights in \eqref{known_weights_unit_res}. 
We refer to this method as MDAM--yr.
Finally, we use 
a bespoke implementation of ``mice'' to impute missing items and randomly sample whole records with replacement as imputations for unit nonresponse. We refer to this method as IH. It does not utilize $\mathcal{A}$. 

\begin{figure}[t]
    \centering
    \includegraphics[width=\linewidth]{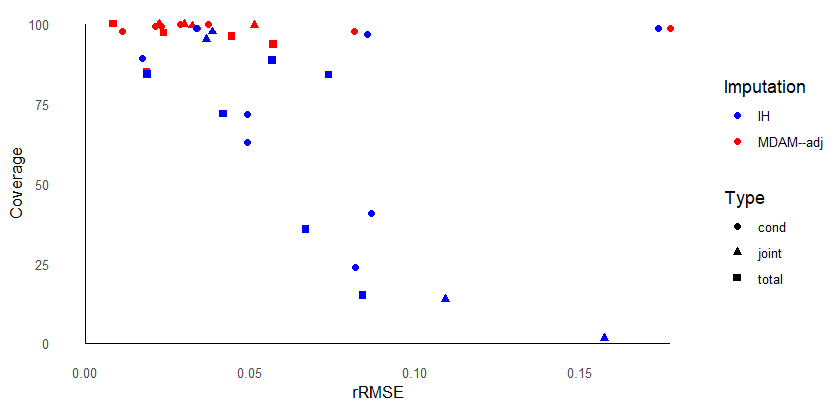}
    \includegraphics[width=\linewidth]{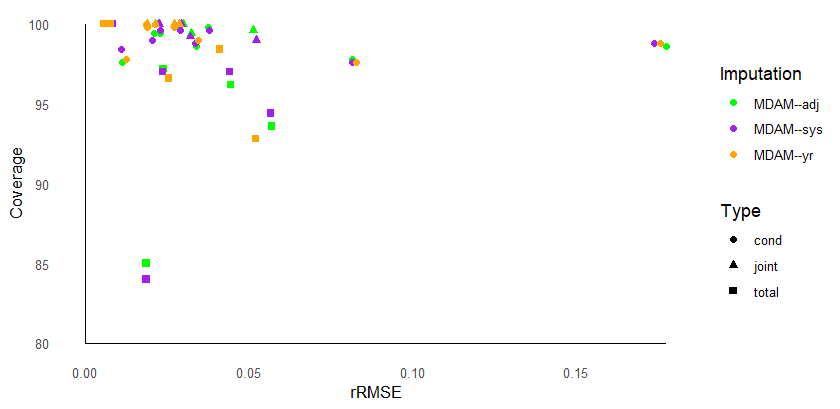}
\caption{Simulated rRMSEs and coverage rates for population totals, two-way conditional probabilities, and joint probabilities. Top panel compares MDAM--adj and IH.  Bottom panel displays MDAM--adj, MDAM--sys, and MDAM--yr.}
\label{fig:sim2}
\end{figure}

Figure \ref{fig:sim2} 
summarizes the relative root mean squared errors (rRMSEs) of the point estimates and empirical coverage rates of 95\% multiple imputation confidence intervals for the  methods across the 500 samples.  Results in tabular form are available at \url{https://github.com/kevinxu47/MDAM}.
MDAM--adj tends to produce lower rRMSEs and higher coverage rates than IH. The coverage rates for the MD-AM models tend to exceed 95\% due to over-estimation of variances.  As discussed by \citet{yang:reiter}, this is because the combining rules of \citet{rubin1987multiple} do not account for using known totals in imputation. 
MDAM--adj and MDAM--sys offer qualitatively similar results. For both, only the interval for $T_{X_3}$ has a lower than nominal coverage rate (around 85\%). 
This rate approximately matches the coverage rate of design-based confidence intervals estimated before introducing any missing data (see the tabular results), suggesting that the lower than nominal coverage is a feature of the complete data estimator, not a consequence of MDAM modeling. 
MDAM--yr performs similarly to MDAM--adj and MDAM--sys.
As evident in the tabular results, MDAM--yr tends to have lower variances, which results from  flattening all weights for unit nonrespondents to a constant.

Overall, the results suggest that the new MD-AM models enable analysts 
to incorporate known marginal totals in multiple imputation when using all sampled units' design weights. 

\bibliographystyle{chicago}
\bibliography{ref.bib}

\end{document}